\documentclass[a4paper]{jpconf}
\usepackage{graphicx}

\newcommand{\ket}{\rangle }
\newcommand{\bra}{\langle }

\newcommand{\ve}{\varepsilon}
\newcommand{\dket}{\ket\ket}
\newcommand{\dbra}{\bra\bra}

\newcommand{\Vect}[1]{\mbox{\boldmath$#1$}}

\begin{document}
\title{All Optical Measurement Proposed for the Photovoltaic Hall Effect}

\author{Takashi Oka, and Hideo Aoki}

\address{Department of Physics, University of Tokyo, Hongo, Tokyo 113-0033, JAPAN}

\ead{oka@cms.phys.s.u-tokyo.ac.jp}

\begin{abstract}
We propose an all optical way to measure the recently 
proposed gphotovoltaic Hall effecth, i.e., a DC Hall effect induced by 
a circularly polarized light in the absence of static magnetic fields.
For this, we have calculated the Faraday rotation angle 
induced by the photovoltaic Hall effect
with the Kubo formula extended for photovoltaic optical response in 
the presence of strong AC electric fields treated with 
the Floquet formalism.  
We also point out the possibility of observing the effect in three-dimensional graphite, and more generally in multi-band systems such as materials described by the dp-model.

\end{abstract}
\section{Introduction}


There are increasing fascinations with new types of Hall effect 
besides the quantum Hall effect in magnetic fields, where a notable example is 
the spin Hall effect in topological insulators.   We have previously 
proposed, as 
another new type of Hall effect, the gphotovoltaic Hall effecth, which should take place when we shine a circularly polarized light to graphene\cite{OkaPHE}. The effect is interesting since this Hall effect occurs in the absence of uniform magnetic fields. Physically, the photovoltaic Hall current originates from the Aharonov-Anandan phase (a non-adiabatic generalization of the geometric (Berry) phase) that an electron acquires during its circular motion around the Dirac points in k-space in an AC field, and bears, in this sense, a topological origin similar to the quantum Hall effect. 
However, unlike the quantum Hall effect, 
the photovoltaic Hall conductivity is in general not quantized. 

We have originally proposed an experimental setup to measure 
the photovoltaic Hall current through the gate electrodes 
attached to the sample.
In the present report we propose a second, all-optical setup, which employs  the Faraday rotation as in the optical Hall effect 
discussed in refs.~\cite{MorimotoOpticalHall,expOpticalHall}, 
which becomes here non-linear optical measurements, since this is a 
``pump-probe" type experiment, where one optically probes the Hall effect 
that is induced in a system pumped by a circularly polarized light 
from a continous laser source.  This effect should grow with the intensity of the applied circularly polarized light.
We also point out the possibility of observing the effect not only in graphene, but also in three-dimensional graphite, and more generally in multi-band systems such as materials described by the $dp$-lattice.

\section{Kubo-formula extended for optical responses in a strong background light}
So let us first propose a method to measure the photovoltaic Hall effect  
with non-linear optical measurements.
The experimental setup is schematically displayed in Fig.\ref{sigma}(a),
where the sample is subject to a strong circularly polarized light
with strength $F$ and frequency (photon energy) $\Omega$.
The strong external AC field changes the properties of the
electronic state and the photo-induced band, 
or the Floquet band to be more precise, can acquire 
a non-trivial topological nature. 
This was discussed in ref.~\cite{OkaPHE}
as an emergence of the photovoltaic Berry's curvature.  
Optical repsonse can be drastically 
change in the presence of the photovoltaic Berry's curvature, 
which we show here to be captured through an optical 
Faraday (or Kerr) 
rotation measurement.  
Our starting point to calculate the optical response is 
the Floquet expression for the 
Kubo formula extendend to incorporate photovoltaic transports \cite{OkaPHE}, 
\begin{eqnarray}
&&\sigma_{ab}(\Vect{A}_{\rm ac})=i\int \frac{d\Vect{k}}{(2\pi)^d}
\sum_{\alpha,\beta\ne\alpha}
\frac{[f_\beta(\Vect{k})-f_\alpha(\Vect{k})]}{\ve_\beta(\Vect{k})-\ve_\alpha(\Vect{k})}
\frac{
\dbra\Phi_\alpha(\Vect{k})|J_b|\Phi_{\beta}(\Vect{k})\dket
\dbra\Phi_\beta(\Vect{k})|J_a|\Phi_{\alpha}(\Vect{k})\dket
}{\ve_\beta(\Vect{k})-\ve_\alpha(\Vect{k})+i\eta}
\end{eqnarray}
with $a,b=x$ or $y$.
The DC expression can be extended to AC responses, 
which reads
\begin{eqnarray}
&&\sigma_{xy}(\omega;\Vect{A}_{\rm ac})=-i\int \frac{d\Vect{k}}{(2\pi)^d}
\sum_{\alpha,\beta\ne\alpha}
\frac{f_\beta(\Vect{k})}{\ve_\beta(\Vect{k})-\ve_\alpha(\Vect{k})}
\label{eq:kuboxy}
\\
&&\hspace{0.8cm}\times
\left[
\frac{
\dbra\Phi_\alpha(\Vect{k})|J_x|\Phi_{\beta}(\Vect{k})\dket
\dbra\Phi_\beta(\Vect{k})|J_y|\Phi_{\alpha}(\Vect{k})\dket
}{\ve_\beta(\Vect{k})-\ve_\alpha(\Vect{k})+\omega+i\eta}-
\frac{
\dbra\Phi_\alpha(\Vect{k})|J_y|\Phi_{\beta}(\Vect{k})\dket
\dbra\Phi_\beta(\Vect{k})|J_x|\Phi_{\alpha}(\Vect{k})\dket
}{\ve_\beta(\Vect{k})-\ve_\alpha(\Vect{k})-\omega+i\eta}
\right],\nonumber
\end{eqnarray}
for the $xy$-component and
\begin{eqnarray}
\mbox{Re}\;\sigma_{xx}(\omega;\Vect{A}_{\rm ac})=\frac{\pi}{\omega}
\int \frac{d\Vect{k}}{(2\pi)^d}
\sum_{\alpha<\beta}
[f_\alpha(\Vect{k})-f_\beta(\Vect{k})]
|\dbra\Phi_\beta(\Vect{k})|J_x|\Phi_{\alpha}(\Vect{k})\dket|^2
\delta(\ve_\beta(\Vect{k})-\ve_\alpha(\Vect{k})-\omega)
\label{eq:kuboxx}
\end{eqnarray}
for the $xx$-component.
Here $|\Phi_\alpha(t)\ket$ is the $\alpha$-th Floquet state, 
the double brakets includes a time average, 
$f_\alpha(\Vect{k})$ the non-equilibrium distribution 
(occupation fraction) of the $\alpha$-th Floquet state, 
and $\eta$ a positive infinitesimal. 
Strictly speaking, we must take into account the 
effect of relaxation (phonons, electrodes, etc.) 
to determine the non-equilibrium distribution function 
as a detailed balanced state with 
photo-absorbtion, photo-emission and the relaxations.
Here we assume for simplicity that the non-equilibrium distribution function
is given by 
\begin{eqnarray}
f_\alpha(\Vect{k})=\sum_i|\dbra\psi_i|\Phi_\alpha(\Vect{k})\dket|^2
f_{\rm FD}(E_i,\mu;\beta_{\rm eff}),
\end{eqnarray}
where $f_{\rm FD}(E_i,\mu;\beta_{\rm eff})=1/[\exp(\beta_{\rm eff}(E_i-\mu))+1]$
is the equilibrium Fermi-Dirac distribution
with an effective inverse temperature $\beta_{\rm eff}$.
This corresponds to employing a sudden approximation.
A proper treatment of the relaxation can be done using the 
Keldysh Green's function as in ref.~\cite{OkaPHE}.

\subsection{Experimental feasibility}
In discussing the experimental feasibility, we can start from the 
relation between the induced conductivities $\sigma_{xx}(\omega), \sigma_{xy}(\omega)$ 
and the Faraday-rotation angle $\Theta_H$ \cite{MorimotoOpticalHall}, 
\begin{eqnarray}
\Theta_H&=&\frac{1}{2}\mbox{arg}\left[
\frac{n_0+n_s+(\sigma_{xx}+i\sigma_{xy})/(c\ve_0)}
{n_0+n_s+(\sigma_{xx}-i\sigma_{xy})/(c\ve_0)}
\right]\\
&=&\frac{1}{(n_0+n_s)c\ve_0}\sigma_{xy}(\omega)
\sim (\sigma_{xy}\;\mbox{in units of }\frac{e^2}{h})\;\times\; 7\;\mbox{mrad}.
\end{eqnarray}
This gives an estimate for the experimental precision
demanded to observe the photovoltaic Hall effect through 
optical measurements.  
For example, in Fig.\ref{sigma} below, we estimate the 
photo-induced response $\sigma_{xy}(\omega;\Vect{A})$ for a single-layered
graphene to be around $0.2$. This corresponds to 
a Faraday-rotation angle of $\Theta_H\sim0.2\times 7=1.4\;\mbox{mrad}$.

\section{Photovoltaic optical response}

\subsection{Single-layer graphene}
\begin{figure}[ht]
\centering 
\includegraphics[width=16.75cm]{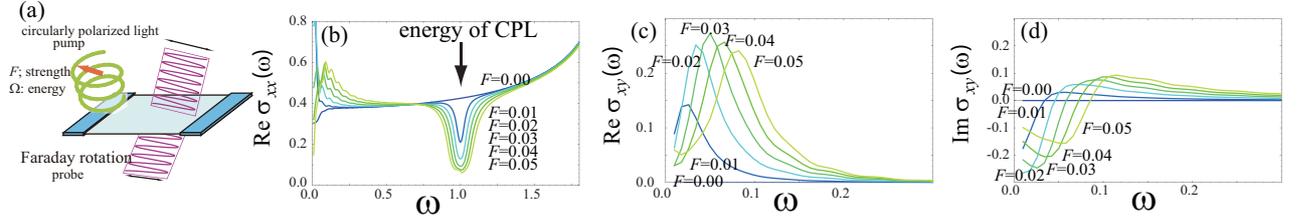}
\caption{
(a)
Schematic, all-optical experimental setup to measure the photovoltaic 
Hall effect via non-linear optical responses.
(b)(c)(d)
Photovoltaic optical response in the
honeycomb lattice (single-layer graphene). 
The optical absorption spectrum 
$\mbox{Re}\sigma_{xx}(\omega;\Vect{A})$ (b) 
and the ``photovoltaic optical Hall coefficient" $\sigma_{xy}(\omega;\Vect{A})$ (c,d) 
are plotted for various values of strength $F$ of the circularly polarized 
light. 
Note the different scales between (b) and (c,d).  
Here the energy of light is fixed to $\Omega=1.0$ 
 in units of the hopping integral ($t \sim 3$ eV for graphene), 
and the effective inverse temperature
$\beta_{\rm eff}=100$. }
\label{sigma}
\end{figure}
Using the extended Kubo formula (eqns.~(\ref{eq:kuboxy}),(\ref{eq:kuboxx})),
we calculate the non-linear optical response in the presence of 
a strong circularly-polarized light in the honeycomb lattice.
The optical absorption spectrum in Fig.~\ref{sigma} (b) 
exhibits the following: 
(i) At the energy of the circularly-polarized light (CPL), there
is a dip in the absorption ($\sigma_{xx}(\omega)$, 
which is an analogue of hole burning. 
(ii) Near the DC limit ($\omega\to 0$), as one increases the 
field strength, the absorption first increases 
due to photo-carriers but then decreases. 
The decrease is due to the opening of a 
photo-induced gap at the Dirac point as discussed in ref.~\cite{OkaPHE}.
The real and imaginary parts of the 
photovoltaic optical {\it Hall} coefficient 
are plotted in Fig.~\ref{sigma} (c,d), 
which grow in a low-frequency region. 
The peak position of the real part 
shifts to higher frequency as the strength of the CPL is 
increased. This is important, since the 
experimental detection is easier for higher photon energies.

\subsection{Multi-layer graphene}
\begin{figure}[ht]
\centering 
\includegraphics[width=14.3cm]{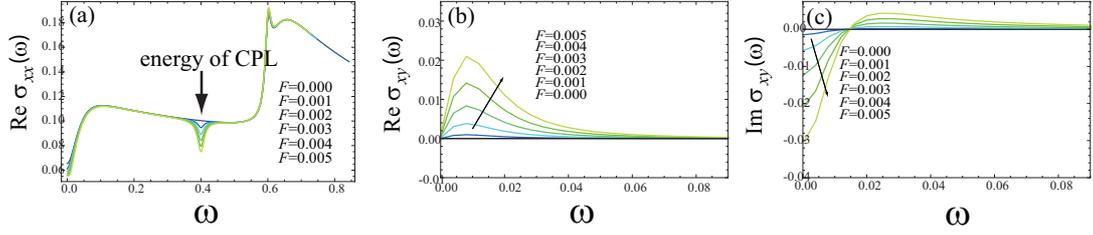}
\caption{
Photovoltaic optical response in a bilayer graphene 
for $\Omega=1.0,\;t_p=0.6$ 
with different scales between (a) and (b,c).  
Note that the input field is an order of magnitude weaker than in 
Fig.\ref{sigma}.
}
\label{sigma_double}
\end{figure}

Photovoltaic Hall effect is not restricted to systems
having a Dirac cone as in the single-layer graphene. 
To show this, we calculate the 
photovoltaic optical response in bilayer graphene (see \cite{GeimRMP}
for notation and refs) described by the Hamiltonian, 
\begin{eqnarray}
H=
\left(
	\begin{array}{cccc}
0 &k_x-ik_y&t_p&0\\
k_x+ik_y& 0 &0&0\\
t_p&0& 0 &k_x+ik_y\\
0&0&k_x-ik_y& 0
\end{array}
\right).
\end{eqnarray}
We plot the photovoltaic optical response in 
Fig.~\ref{sigma_double} for this effective model.
The basic features are similar to those in the monolayer graphene. 
The result suggests the possibility of observing photovoiltaic 
Hall effect in mulilayer graphene and graphite,
since one can block-diagonalize the Hamiltonian for a multilayer system 
into sub-blocks of single and bi-layer graphene components 
as was shown in ref.~\cite{KoshinoAndo07}.

\section{Photovoltaic Berry's curvature in the $dp$-lattice}

\begin{figure}[h]
\centering 
\includegraphics[width=11.5cm]{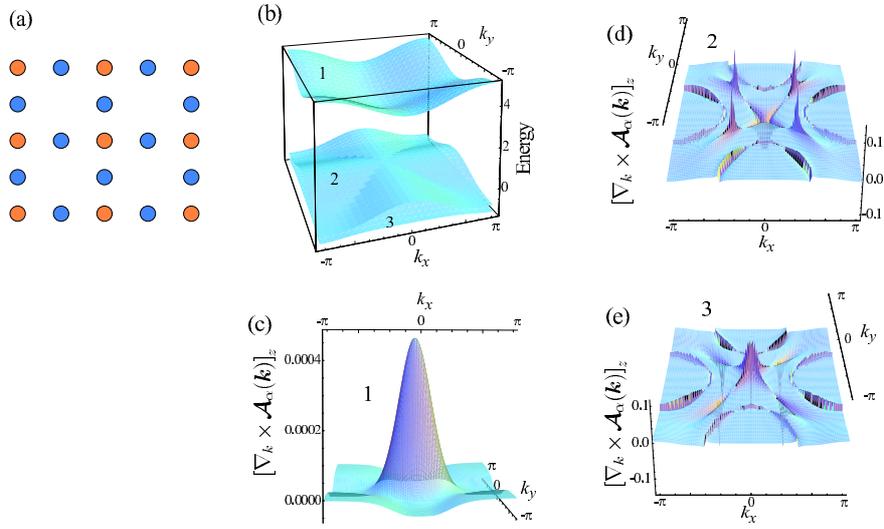}
\caption{
(a) $dp$-lattice.
(b)
The band dispersion in the $dp$ model. 
(c)(d)(e) Photovoltaic Berry's curvature in the three bands 
(top d-band (c), and middle (d), bottom (e) p-bands) for 
$t_{dp}=0.4, \; \ve_d-\ve_p=4,\;F=0.01,\;\Omega=3.$.
}
\label{results2}
\end{figure}

We finally explore the possibility of 
more generally observing the photovoltaic Hall effect in 
materials other than graphene/graphite.  
For a typical multiband model, we consider here the $dp$-lattice, 
which is a well-studied lattice in connection with 
the high-Tc cuprates\cite{ZhangRice}. 
Here we neglect the electron-electron interaction 
to consentrate on the one-particle properties. 
The Hamiltonian is given by
\begin{eqnarray}
H&=&\left(
	\begin{array}{ccc}
\ve_d-\ve_p &-2t\sin\frac{k_x}{2}&-2t\sin\frac{k_y}{2}\\
-2t\sin\frac{k_x}{2}&0&4t_{dp}\sin\frac{k_x}{2}\sin\frac{k_y}{2}\\
-2t\sin\frac{k_y}{2}&4t_{dp}\sin\frac{k_x}{2}\sin\frac{k_y}{2}&0
\end{array}
\right),
\end{eqnarray}
where $\ve_d-\ve_p$  is the potential difference between the $d$ and $p$ bands.  
The photovoltaic DC Hall conductivity
is expressed as Berry's curvature as \cite{OkaPHE,OkaAokiPHBerry}  
\begin{eqnarray}
\sigma_{xy}(\Vect{A}_{\rm ac})=e^2\int \frac{d\Vect{k}}{(2\pi)^d}\sum_\alpha
f_\alpha(\Vect{k})\left[\nabla_{\Vect{k}}\times\Vect{\mathcal{A}}_\alpha(\Vect{k})\right]_z ,
\label{eq:TKNN}
\end{eqnarray}
in terms of a gauge field 
$\Vect{\mathcal{A}}_\alpha(\Vect{k}) \equiv 
-i\dbra\Phi_\alpha(\Vect{k})|\nabla_{\Vect{k}}|\Phi_\alpha(\Vect{k})\dket$.
Note that this expression reduces to the TKNN formula \cite{TKNN}
in the adiabatic limit.

The band dispersion of the $dp$-lattice is shown
in Fig.~\ref{results2} (b)
and the photovoltaic Berry's curvature for each band is plotted
in (c)-(e).
A peak in  Berry's curvature emerges at the center of the
Brilliouin zone in the $d$-band, while the curvatue has 
complex structures in the $p$-bands.
This is because the two $p$-bands intersect with each other, 
for which the ac field induces a band mixing, 
while the $d$-band is separated from the 
$p$-bands with the separation larger than the 
photon energy $\Omega$ considered in the figure.

\begin{figure}[htb]
\centering 
\includegraphics[width=13.cm]{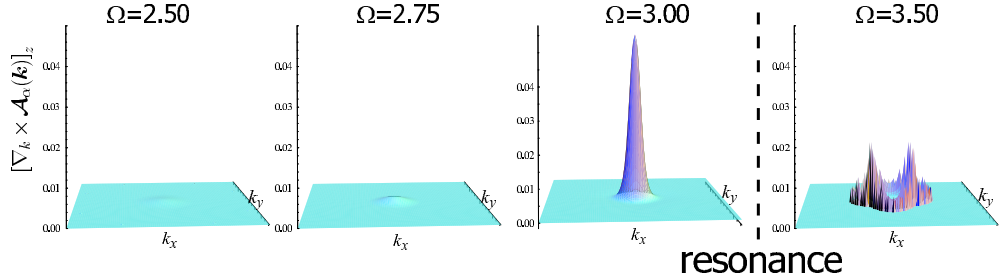}
\caption{Photovoltaic Berry's curvature in the $d$-band
of the $dp$-lattice for severa values of thel photon energy $\Omega$ 
for $t_{dp}=0.4, \;\ve_d-\ve_p=4,\;F=0.1$.
}
\label{berry}
\end{figure}

Now, a natural question is: Is there a 
difference between a system with 
a band gap as in the $dp$-lattice when 
compared to a system with a massless Dirac cone as in graphene? 
In order to clarify this, we  have 
calculated the photovoltaic Berry's curvature for 
several values of $\Omega$ in Fig.~\ref{berry}.
When $\Omega$ is smaller than the gap, 
the photovoltaic Berry curvature increases as 
the value of $\Omega$ approaches the band gap.   
When $\Omega$ exceeds the 
gap, when a direct photo-transition 
becomes possible, the 
Berry's curvature changes into a concentric 
form with a negative part (not apparent in the figure). 
In the massless Dirac case, 
both the cone-like contribution and 
the concentric form coexist\cite{OkaPHE,OkaAokiPHBerry},
while in the $dp$-lattice they appear separately. 
Another important obervation is that 
the magnitude of the photovoltaic Hall effect becomes large in 
multiband systems when 
$\Omega$ is slightly below the band gap, whereas in the massless 
Dirac case small $\Omega$ is better. 

To conclude, we have studied the possibility of observing the 
photovoltaic Hall
effect in an all-optical fashion. We have also shown that 
the effect is not restricted to 
lattices with a massless Dirac cone but is a universal phenomenon 
in multi-band systems. 
HA has been supported in part by a Grant-in-Aid for Scientific Research 
No.20340098 from JSPS, 
TO by a Grant-in-Aid for Young Scientists (B) 
from MEXT
and by Scientific Research on Priority Area
``New Frontier of Materials Science Opened by Molecular Degrees of Freedom".


\section*{References}
\begin{thebibliography}{9}

\bibitem{OkaPHE} T. Oka, and H. Aoki, {\it Phys. Rev.} {\bf B  79}, 081406 (R) (2009); ibid {\bf 79}, 169901(E) (2009). 
\bibitem{MorimotoOpticalHall} T. Morimoto, Y. Hatsugai, and H. Aoki, {\it Phys. Rev. Lett.} {\bf 103}, 116803 (2009).
\bibitem{expOpticalHall} Y. Ikebe, T. Morimoto, R. Masutomi, T. Okamoto, H. Aoki, and R. Shimano, {\it Phys. Rev. Lett.} {\bf 104}, 256802 (2010).

\bibitem{GeimRMP}
A. H. Castro Neto, F. Guinea, N. M. R. Peres, K. S. Novoselov, and A. K. Geim,
{\it Rev. Mod. Phys.} {\bf 81}, 109 (2009).

\bibitem{KoshinoAndo07}
M. Koshino and T. Ando, Phys. Rev. B {\bf 76}, 085425 (2007). 

\bibitem{OkaAokiPHBerry} T. Oka and H. Aoki, 
{\it J. Phys.: Conf. Ser.} {\bf 200}, 062017 (2010).

\bibitem{ZhangRice} F. C. Zhang and T. M. Rice,
{\it Phys. Rev.} {\bf B 37}, 3759 (1988).

\bibitem{TKNN} D. J. Thouless, M. Kohmoto, M. P. Nightingale, and M. den Nijs,
{\it Phys. Rev. Lett.} {\bf 49}, 405 (1982).

\end{thebibliography}

\end{document}